\documentclass[journal,10pt]{IEEEtran}

\usepackage[cmex10]{amsmath}
\usepackage[utf8]{inputenc}
\usepackage[pdftex]{graphicx}
\usepackage{epstopdf}
\usepackage{array}
\usepackage{floatrow}
\usepackage{caption}
\usepackage{subcaption}
\usepackage{url}
\usepackage{balance}
\usepackage{enumerate} 
\usepackage{amsbsy}
\usepackage[normalem]{ulem}
\usepackage{textcomp}
\usepackage{siunitx}
\usepackage{balance}
\usepackage{placeins}
\usepackage{tabularx, ragged2e}
\usepackage{listings}
\usepackage[dvipsnames]{xcolor}
\usepackage{svg}
\usepackage{epstopdf}
\usepackage{amssymb}   
\usepackage{pifont}    
\usepackage{fontawesome5}
\usepackage{multirow}
\usepackage{booktabs}

\def\BibTeX{{\rm B\kern-.05em{\sc i\kern-.025em b}\kern-.08em
    T\kern-.1667em\lower.7ex\hbox{E}\kern-.125emX}}
    
\usepackage{xspace}

\DeclareTextFontCommand{\textcomputer}{\fontfamily{cmr}\selectfont}

\begin{document}

\title{From Isolated Protocols to Integrated Pipelines: Benchmarking IoT Communication in the MASA Living Lab}


\author{\IEEEauthorblockN{
Antonio Solida$^1$,
Enrico Rossini$^2$,
Marcello Pietri$^2$,
Marco Picone$^2$,
Carlo Augusto Grazia$^1$\\}
\IEEEauthorblockA{
\textit{$^1$Department of Engineering ``Enzo Ferrari'',\\ $^2$Department of Science and Methods for Engineering,}\\
\textit{University of Modena and Reggio Emilia}\\
\{name.surname\}@unimore.it
}
\\
}

\maketitle
\pagestyle{empty}
\thispagestyle{empty}


\begin{abstract}
The performance of real-time IoT communication systems depends not only on individual transport protocols, serialization formats, and data-processing stages, but also on how these components interact within the complete communication pipeline. Selecting an appropriate communication stack is therefore application-dependent and requires considering processing capabilities, network characteristics, communication patterns, and reliability requirements. This paper provides an experimental, evidence-based analysis of these interactions through a controlled benchmark of integrated communication pipelines deployed in the MASA living lab. We jointly evaluate local processing overhead, end-to-end latency, jitter, throughput, and delivery reliability across different combinations of transport, messaging, serialization, and compression mechanisms. Results show that component-level processing costs do not directly predict end-to-end performance. In particular, differences of only tens of microseconds in serialization or compression costs can coexist with several-millisecond differences in end-to-end latency, while middleware buffering and transport behavior substantially affect timing performance. The results demonstrate that communication-stack components should be evaluated as integrated pipelines and under representative operating loads rather than ranked independently.
\end{abstract}


\section{Introduction}
\label{sec_intro}
Cyber-Physical Systems (CPS) rely on timely data exchange between distributed devices, supported by IoT and Industrial IoT technologies in domains ranging from smart cities to industrial environments. Meeting latency, throughput, and reliability requirements depends on the combined behavior of transport protocols, messaging mechanisms, serialization, and compression. Their individual costs may not predict overall performance: faster serialization or smaller payloads can interact with transport and middleware buffering in ways that alter end-to-end (E2E) latency.

Existing benchmarks typically evaluate these components separately, measuring transport performance, MQTT behavior under load, or local serialization and compression costs. Such evaluations do not establish whether component-level rankings hold when these technologies are combined into a communication pipeline. We therefore investigate three questions: can local processing times predict communication E2E latency; how do serialization, compression, and middleware interact; and how does offered load affect latency, jitter, throughput, and delivery reliability?

\begin{figure}[h!]
\centering
\includegraphics[width=0.70\textwidth]{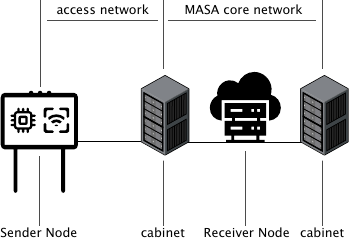}
\caption{MASA: real living-lab network.}
\label{fig:masa}
\end{figure}
\vspace{-12pt}
This paper presents a controlled benchmark of TCP and UDP transport, ZeroMQ and MQTT messaging (Table \ref{tab:communication_technologies}), CBOR and Protocol Buffers serialization (Table \ref{tab:data_processing_formats}), and Zstandard compression, evaluated individually and in combination. These technologies form a selected subset of the MASA software ecosystem, covering reliable and best-effort transport, different messaging mechanisms, alternative binary serialization formats, and lossless compression. We measure communication E2E delay separately from local processing costs, alongside jitter, throughput, and delivery reliability, to examine whether processing-cost differences translate into communication performance or are affected by transport and middleware interactions.

All experiments run on hardware deployed in the MASA (Modena Automotive Smart Area) living lab, using its high-speed fiber-optic network \cite{10770429} (Figure \ref{fig:masa}). This controlled, low-loss setting reduces variability from lower network layers, helping isolate the effects of the evaluated pipeline components.

The paper is organized as follows.
Section~\ref{sec_related1} provides an overview of related works.
Section~\ref{methodology} presents the tests, Section~\ref{experimental_results} analyzes the results, and
Section~\ref{sec_conclusions} concludes the article.

\begin{table}[t]
\caption{Communication technologies evaluated in the benchmark.}
\label{tab:communication_technologies}
\centering
\footnotesize
\begin{tabular}{@{}l p{2.7cm} p{4cm}@{}}
\toprule
\textbf{Acronym} & \textbf{Full Name} & \textbf{Description} \\
\midrule
TCP  & Transmission Control Protocol
     & Reliable connection-oriented transport \\

UDP  & User Datagram Protocol
     & Best-effort connectionless transport \\

ZMQ  & ZeroMQ
     & Messaging middleware \\

MQTT & Message Queuing Telemetry Transport
     & Publish/subscribe messaging protocol \\
\bottomrule
\end{tabular}
\end{table}

\begin{table}[t]
\caption{Data processing formats evaluated in the benchmark.}
\label{tab:data_processing_formats}
\centering
\footnotesize
\begin{tabular}{@{}l p{3.2cm} p{3.6cm}@{}}
\toprule
\textbf{Acronym} & \textbf{Full Name} & \textbf{Description} \\
\midrule
CBOR     & Concise Binary Object Representation
         & Schema-less binary serialization \\

Protobuf & Protocol Buffers
         & Schema-based binary serialization \\

Zstd     & Zstandard
         & Lossless data compression \\
\bottomrule
\end{tabular}
\end{table}

\section{Related Work}
\label{sec_related1}
Researchers have widely studied communication performance in IoT systems by measuring latency, throughput, packet loss, and protocol overhead. Moraes et al. compared IoT protocols in terms of performance and reliability \cite{8914552}, while Handosa et al. examined MQTT under different operating conditions \cite{8248196}. Malhi et al. compared MQTT and CoAP and showed that their relative performance depends on workload and network conditions \cite{10410639}. For messaging middleware, Kang et al. evaluated DDS, MQTT, and ZeroMQ under periodic, sporadic, and high-frequency pub/sub traffic, reporting broader QoS support and generally more reliable performance for DDS, with results depending on traffic patterns \cite{kang2020study}. Proos et al. jointly assessed messaging protocols and serialization formats for vehicle-to-cloud digital twins, finding that UDP-based communication reduces latency and overhead at the expense of reliability guarantees \cite{proos2020performance}. Other work proposed a standardized MQTT broker benchmarking methodology \cite{gheorghe2020performance} and extended MQTT architectures with message manipulation, filtering, and re-routing to offload subscriber computation and optimize processing and traffic \cite{11096145}.

Data representation and buffering also affect communication performance. CBOR has been adopted as a compact representation for constrained IoT environments \cite{8661513}. Jackson et al. compared thirteen serialization formats across eight datasets, measuring serialization time, deserialization time, and payload size \cite{10734127}. They found protocol-based formats fastest, followed by schema-less binary and text-based formats, but did not evaluate how these differences affect E2E latency with transport and middleware. Experimental work has also shown how buffer configuration affects IoT communication performance \cite{9803729}.

These studies examine individual components or selected combinations. Our benchmark considers transport, messaging middleware, serialization, compression, and application-layer reliability within a common pipeline. By measuring local processing costs separately from communication E2E delay, we assess whether component-level performance translates into communication performance, including the effects of middleware buffering and scheduling.

\section{Methodology}
\label{methodology}
All experiments use a custom benchmarking application running on two separate Linux devices. Dedicated communication classes implement a common Sender/Receiver interface, preserving the measurement and data-processing pipeline across configurations. Payload, transmission rate, duration, and protocol-specific parameters remain fixed within each comparison.

Raw TCP uses a 32-bit length prefix, with the receiver buffering bytes until a complete message is reconstructed. MQTT uses Eclipse Paho C++ over plain TCP with a clean session. Synchronous publication waits for the configured QoS exchange to complete (PUBACK for QoS 1; PUBREC/PUBREL/PUBCOMP for QoS 2), limiting publication rates at higher QoS levels. The receiver subscribes at the publication QoS and forwards payloads unmodified through an asynchronous callback. ZeroMQ supports PUSH/PULL and PUB/SUB over TCP: the receiver binds in PUSH/PULL, while the publisher binds in PUB/SUB and subscribers connect with a subscription filter. The receiver polls non-blockingly, sleeping for approximately 100~$\mu$s between empty polls, and reassembles multipart PUB/SUB frames before timestamping. This implementation-specific polling interval can add approximately 100~$\mu$s of delay and is not an intrinsic ZeroMQ overhead.

Unless otherwise specified, each run comprises a 5-second warm-up and a 30-second measurement window. The warm-up accommodates PUB/SUB slow-joiner transients. Each scenario is repeated five times; repeatability is assessed through per-run median E2E latency, with variability summarized by the minimum and maximum medians. No systematic drift is observed.

Experiments use the MASA living lab's high-speed fiber-optic network. This controlled, low-loss setting reduces lower-layer variability, providing a baseline for subsequent evaluation over wireless and constrained networks. Communication E2E latency spans the timestamps acquired immediately before sender transmission and immediately after receiver reception. Serialization, compression, decompression, and deserialization fall outside this interval and are timed separately.

Before each run, both nodes synchronize via SNTP (RFC 4330) against a local MASA time server. Each node performs $N=8$ queries and selects the offset associated with the minimum round-trip time,
$\mathrm{RTT}=(T_4-T_1)-(T_3-T_2)$, to reduce the influence of variable network delay. Per-run synchronization uncertainty is estimated as:
\begin{equation}
\mathrm{ntp\_sync\_uncertainty} =
\frac{\mathrm{RTT}_{\mathrm{sender}}+\mathrm{RTT}_{\mathrm{receiver}}}{2}.
\end{equation}
Although minimum-RTT selection does not eliminate path asymmetry, the estimated uncertainty over the local fiber network remains well below the observed millisecond-scale latency differences. Absolute sub-microsecond precision requires PTP\footnote{https://standards.ieee.org/ieee/1588/4355/} with hardware timestamping. Corrected timestamps provide E2E latency measurements; throughput, jitter, local processing costs, and reliability are measured separately, with reliability expressed as delivery ratio and loss rate.

\subsection{Statistical analysis}
We repeated each experimental scenario five times. For each run, we discarded the first 5 s as warm-up and characterized the E2E latency distribution using the median, standard deviation, 95th and 99th percentiles, and the minimum and maximum observed values. Jitter was computed as $J_i = |L_i-L_{i-1}|$, where $L_i$ is the E2E latency of the $i$-th received message. The values reported in Table~\ref{tab:full-results} summarize the five repetitions, while the Min--Max column reports the overall range of packet-level E2E latency observed across the experiments.

\begin{table*}[t]
\centering
\caption{Complete summary of experimental results.}
\label{tab:full-results}
\footnotesize
\resizebox{\textwidth}{!}{%
\begin{tabular}{@{}cllcccccc@{}}
\toprule
\textbf{\#} & \textbf{Test} & \textbf{Configuration} &
\textbf{Median [ms]} & \textbf{Min--Max [ms]} &
\textbf{$\sigma$ [ms]} & \textbf{Jitter [$\mu$s]} &
\textbf{p95 [ms]} & \textbf{p99 [ms]} \\
\midrule

1  & Baseline & TCP
   & 1.86 & [1.43--6.04] & 0.08 & 25 & 1.92 & 1.96 \\
2  & Baseline & UDP
   & 1.40 & [1.37--3.14] & 0.03 & 15 & 1.43 & 1.46 \\
3  & Baseline & ZMQ
   & 7.35 & [7.22--13.59] & 0.20 & 150 & 7.47 & 8.02 \\

\midrule

4  & Serialization & TCP + CBOR
   & 9.03 & [8.88--9.49] & 0.05 & 45 & 9.10 & 9.14 \\
5  & Serialization & TCP + Protobuf
   & 1.88 & [1.75--2.30] & 0.05 & 45 & 1.96 & 1.99 \\
6  & Serialization & ZMQ + Protobuf
   & 6.78 & [6.61--8.69] & 0.09 & 75 & 6.90 & 6.96 \\

\midrule

7  & Compression & TCP + Zstd + Protobuf
   & 2.77 & [2.56--3.08] & 0.05 & 40 & 2.86 & 2.91 \\
8  & Compression & ZMQ + Zstd + Protobuf
   & 2.41 & [2.25--3.01] & 0.07 & 75 & 2.51 & 2.55 \\

\midrule

9  & Protocol comparison & TCP
   & 1.99 & [1.79--2.60] & 0.08 & 60 & 2.10 & 2.20 \\
10 & Protocol comparison & UDP
   & 1.28 & [1.18--1.79] & 0.06 & 30 & 1.42 & 1.52 \\
11 & Protocol comparison & ZMQ
   & 1.30 & [1.08--1.74] & 0.09 & 80 & 1.46 & 1.53 \\
12 & Protocol comparison & MQTT/TCP QoS 0
   & 1.29 & [1.11--1.63] & 0.06 & 60 & 1.38 & 1.47 \\
13 & Protocol comparison & MQTT/TCP QoS 1
   & 1.24 & [1.05--1.98] & 0.07 & 60 & 1.35 & 1.41 \\

\midrule

14 & High-rate & TCP
   & 4.32 & [4.02--8.10] & 0.19 & 190 & 4.53 & 4.61 \\
15 & High-rate & UDP
   & 2.69 & [2.63--7.93] & 0.13 & 25 & 2.74 & 2.88 \\

\midrule

16 & Reliability & UDP
   & 11.36 & [11.29--13.04] & 0.04 & 40 & 11.41 & 11.44 \\
17 & Reliability & MQTT/TCP QoS 2
   & 0.35 & [0.13--7.35] & 0.29 & 63 & 0.51 & 1.05 \\

\bottomrule
\end{tabular}%
}
\end{table*}
\section{Experimental Tests and Results}
\label{experimental_results}
\subsection{Baseline tests}
The first set of experiments isolates the transport layer by exchanging raw JSON batches at a fixed rate of approximately 1000 msg/s (512 kB/s) across three configurations: raw TCP, raw UDP, and ZeroMQ PUSH/PULL over TCP. All configurations were evaluated under the same load for 30 s with a warm-up phase of 5 s, allowing comparison of delay, jitter, and throughput under identical traffic conditions.

\begin{figure}
        \centering
        \includegraphics[width=\textwidth]{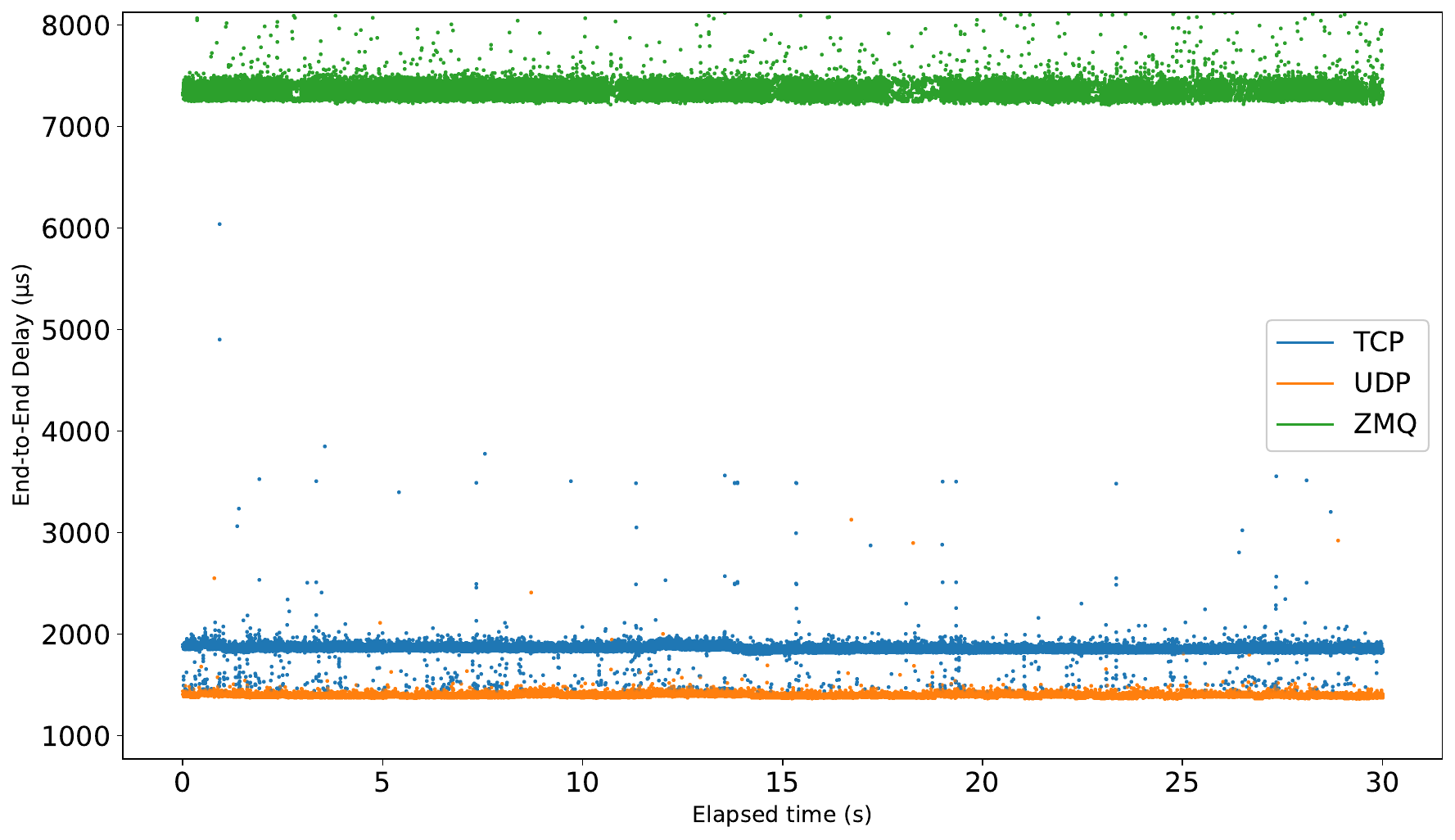} 
        \caption{End-to-end latency.}
        \label{fig:baseline_e2e_latency}
    \end{figure}

\begin{figure}
        \centering
        \includegraphics[width=\textwidth]{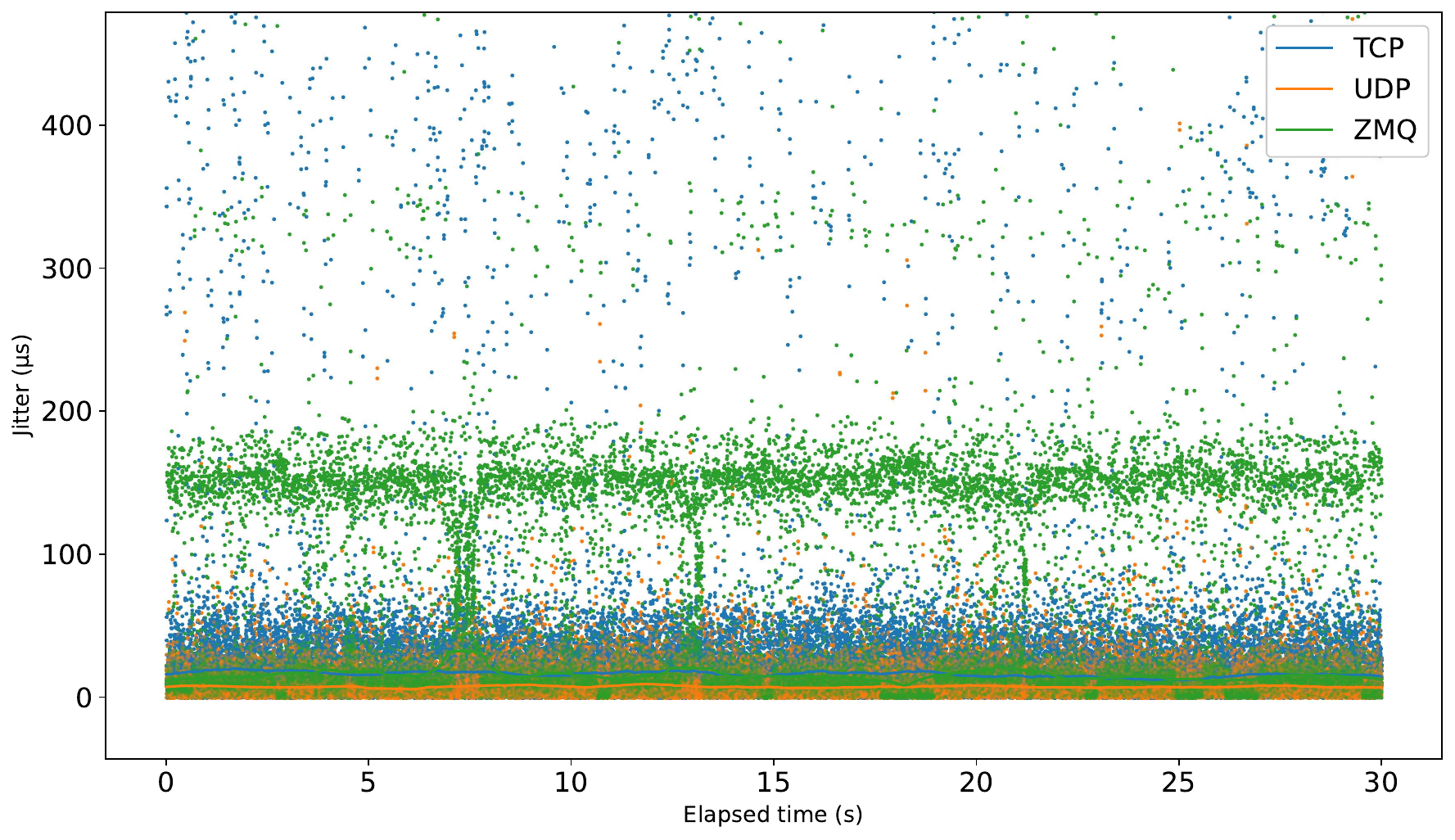} 
        \caption{End-to-end jitter.}
        \label{fig:baseline_e2e_jitter}
    \end{figure}

The three configurations exhibit clearly separated delay levels throughout the experiment. UDP achieves the lowest E2E delay, with a median of approximately 1.4 ms and a narrow distribution. TCP increases the median to about 1.9 ms, with only a small number of samples reaching several milliseconds. ZeroMQ over TCP shows a substantially higher delay, with values concentrated around 7.3–7.5 ms. This is about four times higher than raw TCP and five times higher than UDP. The difference remains stable throughout the test, although short gaps are visible in the ZeroMQ delay trace, suggesting temporary buffering or scheduling effects within the middleware (Figure \ref{fig:baseline_e2e_latency}). Jitter follows the same trend: TCP and UDP remain close to zero for most of the experiment, whereas ZeroMQ stays at about 150 µs without increasing over time. This suggests the additional variability mainly comes from middleware processing and buffering, not an unstable communication channel (Figure \ref{fig:baseline_e2e_jitter}). Throughput remains comparable across the three configurations, staying around 510–514 kB/s and approximately 1000 msg/s throughout the experiment. Therefore, the higher delay and jitter observed with ZeroMQ cannot be attributed to a higher offered load or to throughput limitations, but are consistent with additional processing, buffering, and scheduling effects introduced by the middleware.

Overall, the baseline experiments provide the reference for the reliability-latency trade-off between UDP and TCP, while showing that ZeroMQ introduces a significant additional latency and jitter under otherwise identical conditions. This difference is already present before introducing serialization and compression into the communication pipeline.


\subsection{Serialization}
The second set of experiments compares CBOR and Protobuf over raw TCP, and Protobuf over ZeroMQ, measuring local processing costs separately from communication E2E latency.

CBOR serialization requires approximately 70--75~$\mu$s/packet, compared with 16--20~$\mu$s for Protobuf, a roughly fourfold reduction. Using ZeroMQ leaves Protobuf serialization time essentially unchanged. The transport stage takes approximately 3.5~$\mu$s for CBOR, 4.5--4.7~$\mu$s for Protobuf over TCP, and 6~$\mu$s for Protobuf over ZeroMQ; deserialization remains below 1~$\mu$s for most packets.

\begin{figure}[h!]
        \includegraphics[width=\textwidth]{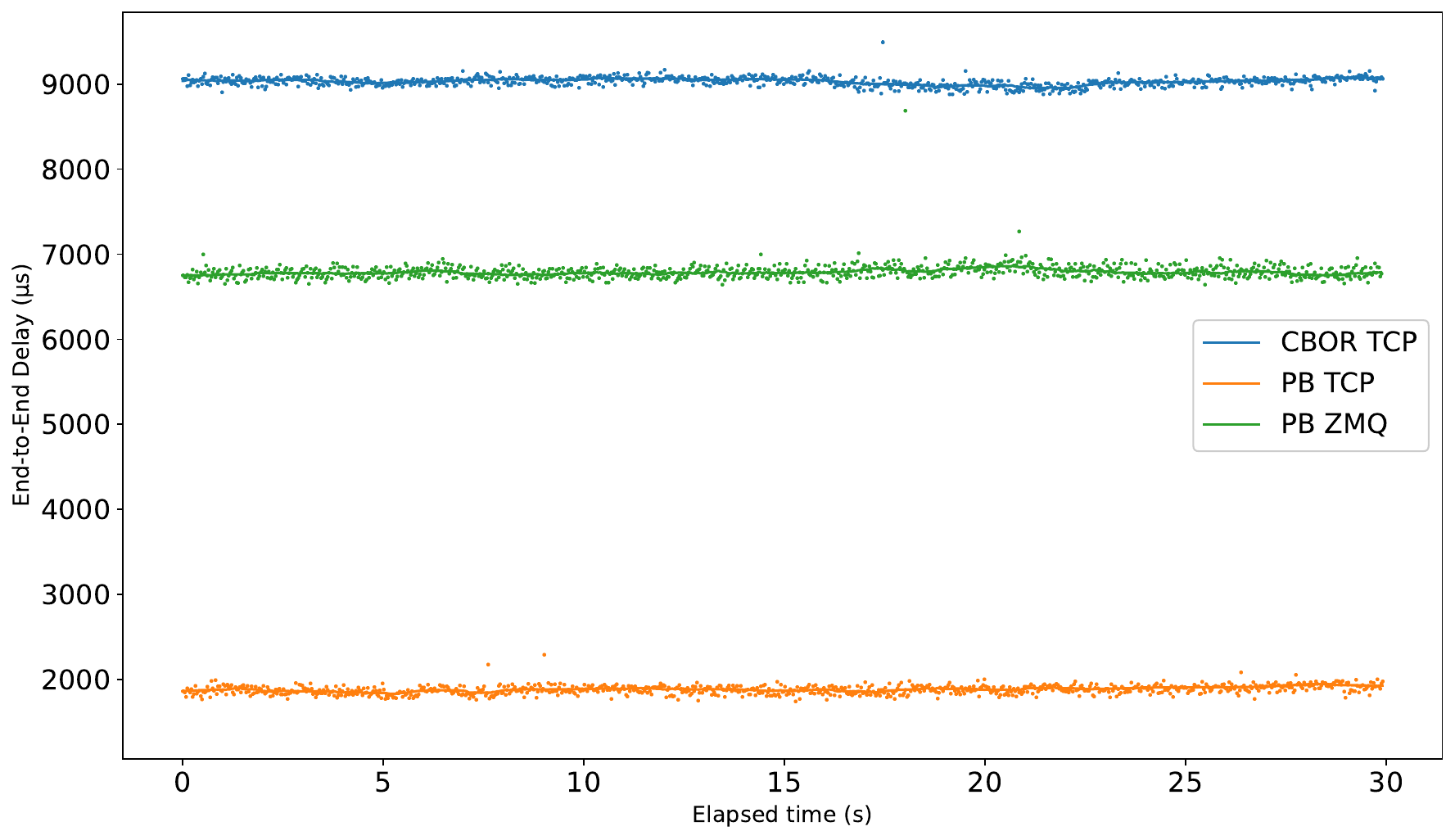} 
        \caption{Serialization End-to-end Latency.}
        \label{fig:seriaz_e2e_lat}
\end{figure}

\begin{figure}[h!]
        \includegraphics[width=\textwidth]{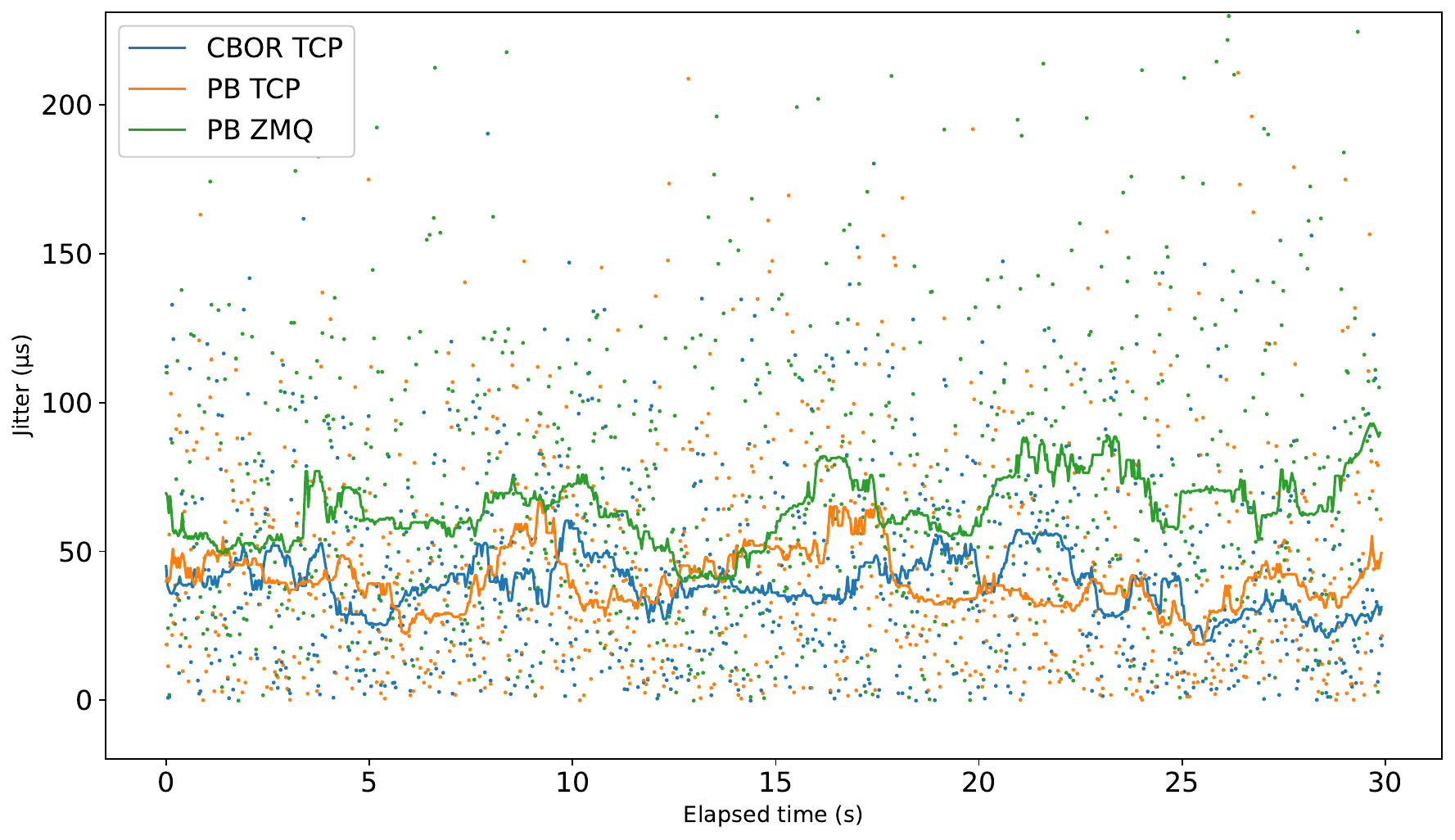} 
        \caption{Serialization End-to-end Jitter.}
        \label{fig:seriaz_e2e_jit}
\end{figure}

Median E2E delays are approximately 9.0, 1.88, and 6.78~ms, respectively (Figure \ref{fig:seriaz_e2e_lat}). The 50--60~$\mu$s difference in serialization cost between CBOR and Protobuf cannot explain their approximately 7~ms communication latency gap, particularly since serialization is measured outside the E2E interval. Likewise, ZeroMQ increases E2E latency by approximately 3.6 times despite using the same Protobuf representation. Its jitter is also higher: 60--90~$\mu$s versus 30--60~$\mu$s for the other configurations (Figure \ref{fig:seriaz_e2e_jit}). These differences are consistent with middleware buffering, scheduling, and transport interactions, although the measurements do not isolate their individual contributions.

Protobuf therefore reduces local serialization cost, but communication latency must be evaluated separately for each pipeline configuration.

\subsection{Compression}
\begin{figure}
        \includegraphics[width=\textwidth]{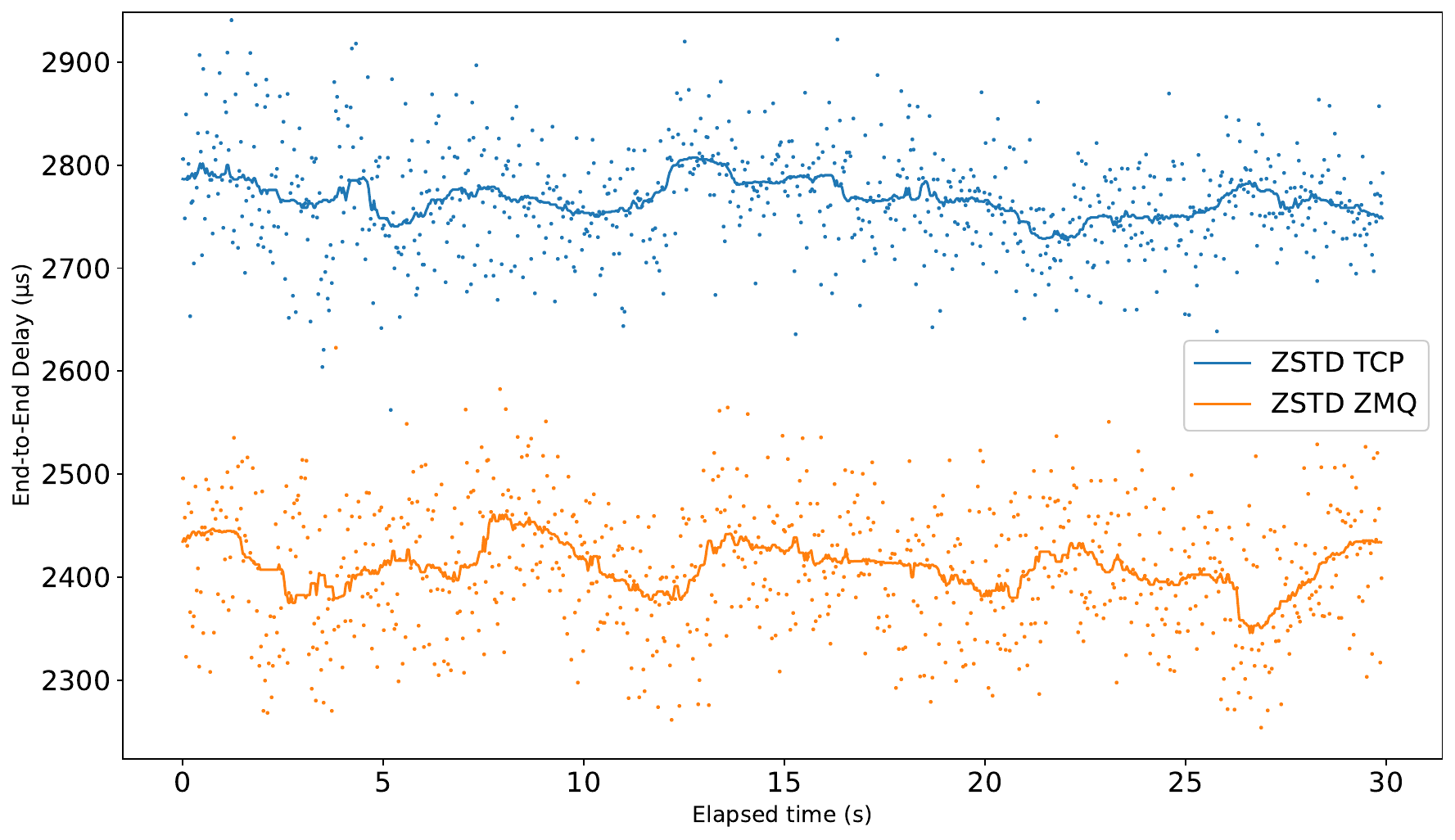} 
        \caption{Compression End-to-end Latency.}
        \label{fig:comp_e2e_lat}
\end{figure}

The compression experiments evaluate the trade-off between Zstd processing overhead and reduced transmitted data size. One configuration combines TCP, Protobuf, and Zstd, while the other applies the same compression stage within a ZeroMQ-based pipeline. The comparison therefore captures both the direct computational cost of compression and its impact on the subsequent transport and queueing stages (Figure \ref{fig:comp_e2e_lat}).

The TCP-based configuration exhibits an initial compression cost of approximately 100–120 µs during the first 7 s of the experiment, followed by a steady-state value of 45–55 µs/packet. The ZeroMQ-based pipeline shows no comparable initialization phase and remains around 60–65 µs. A similar difference appears during decompression: the ZeroMQ pipeline stays between 85 and 100 µs, with occasional peaks up to 130 µs, whereas the TCP-based configuration decreases from about 70 µs to 45–50 µs over the experiment.

Despite the higher steady-state processing cost, the ZeroMQ-based pipeline achieves a lower E2E delay, remaining between 2350 and 2460 µs, compared with 2750–2820 µs for the TCP-based configuration. Since both configurations transmit equivalently compressed payloads, the lower ZeroMQ latency cannot be attributed to payload size or local processing cost, indicating different transport and buffering behavior. The lower latency, however, comes with higher variability: jitter remains between approximately 60 and 90 µs in the ZeroMQ-based configuration, compared with 25–55 µs in the TCP-based one. Throughput is comparable in both cases, remaining within 4800–5600 B/s and 27–32 msg/s. The observed latency difference therefore appears mainly related to processing and buffering behavior rather than differences in the achieved data rate.

\subsection{Protocol comparison}
The main protocol comparison evaluates TCP, UDP, ZeroMQ, MQTT QoS 0, and MQTT QoS 1 under the same payload, publication rate, and processing pipeline. The experiments use a fixed rate of 30 Hz, providing a common workload for comparing transport-level reliability, middleware processing, and application-layer acknowledgment mechanisms.

\begin{figure}[h!]
        \includegraphics[width=\textwidth]{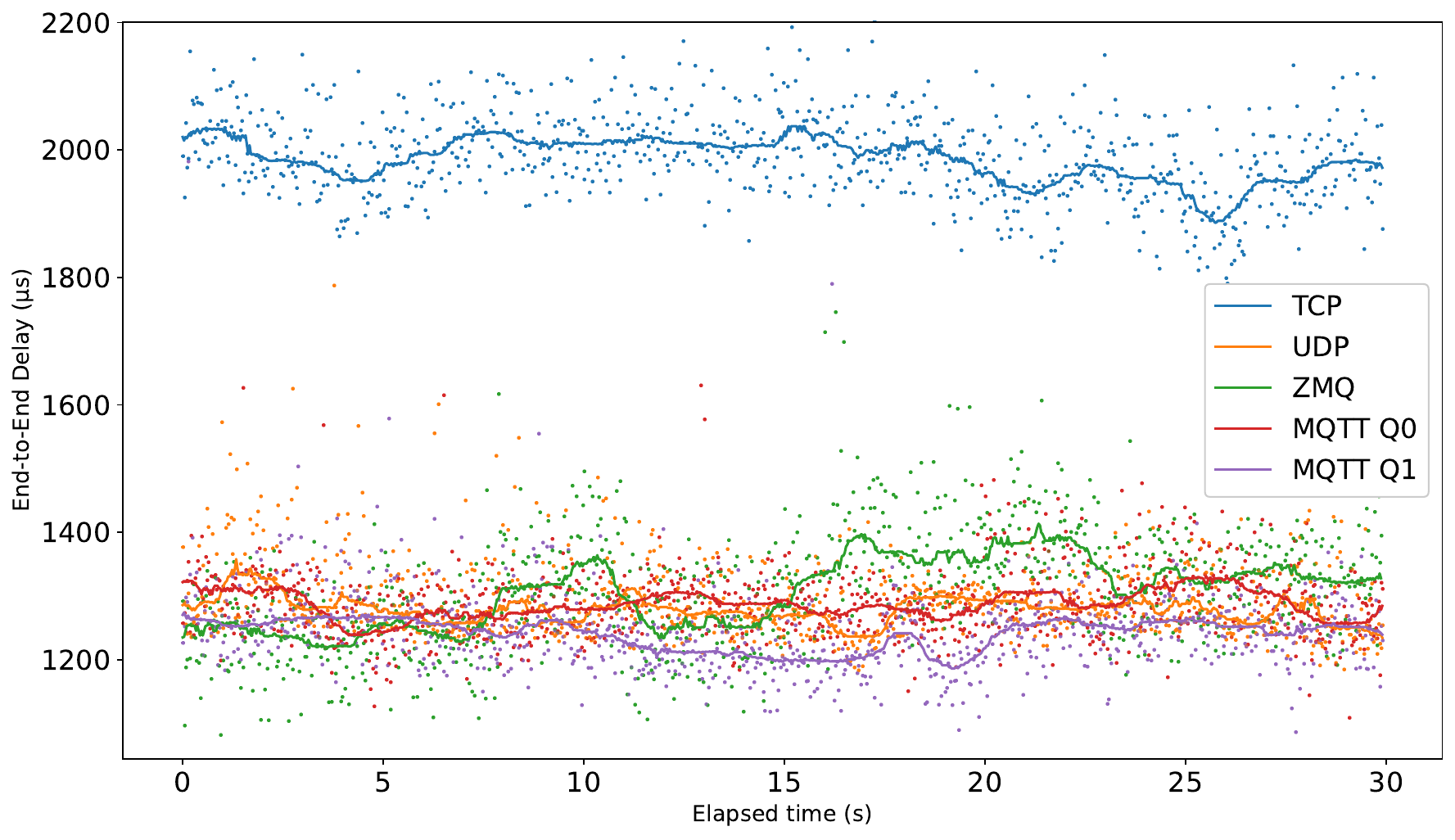} 
        \caption{Protocols comparison End-to-end Latency.}
        \label{fig:cfr_e2e_lat}
\end{figure}

Figure \ref{fig:cfr_e2e_lat} reports the E2E delay measured under the five configurations. TCP exhibits the highest median delay, at approximately 2000 µs, while the other four configurations remain between 1100 and 1450 µs. The difference is therefore not negligible and indicates a measurable cost associated with the TCP communication path under the considered workload. ZeroMQ shows a gradual increase in delay, from about 1250 µs during the first 10 s to 1350–1400 µs after 20 s, suggesting accumulating internal queueing over time. UDP, MQTT QoS 0, and MQTT QoS 1 do not show a comparable drift. The jitter measurements show a similar ordering after the initial 5 s warm-up period. UDP provides the lowest and most stable jitter, remaining between approximately 20 and 40 µs. ZeroMQ exhibits the highest values, typically between 60 and 100 µs, consistent with the delay trend. TCP and the two MQTT configurations occupy an intermediate range of approximately 50–70 µs. Throughput remains within 4800–5600 B/s, corresponding to 27–32 msg/s, for all configurations. Thus, the measured differences in latency and jitter are not due to throughput degradation, but to the processing and queueing behavior of the individual protocols.

\subsection{High-Rate Tests}

\begin{figure}[h!]
        \includegraphics[width=\textwidth]{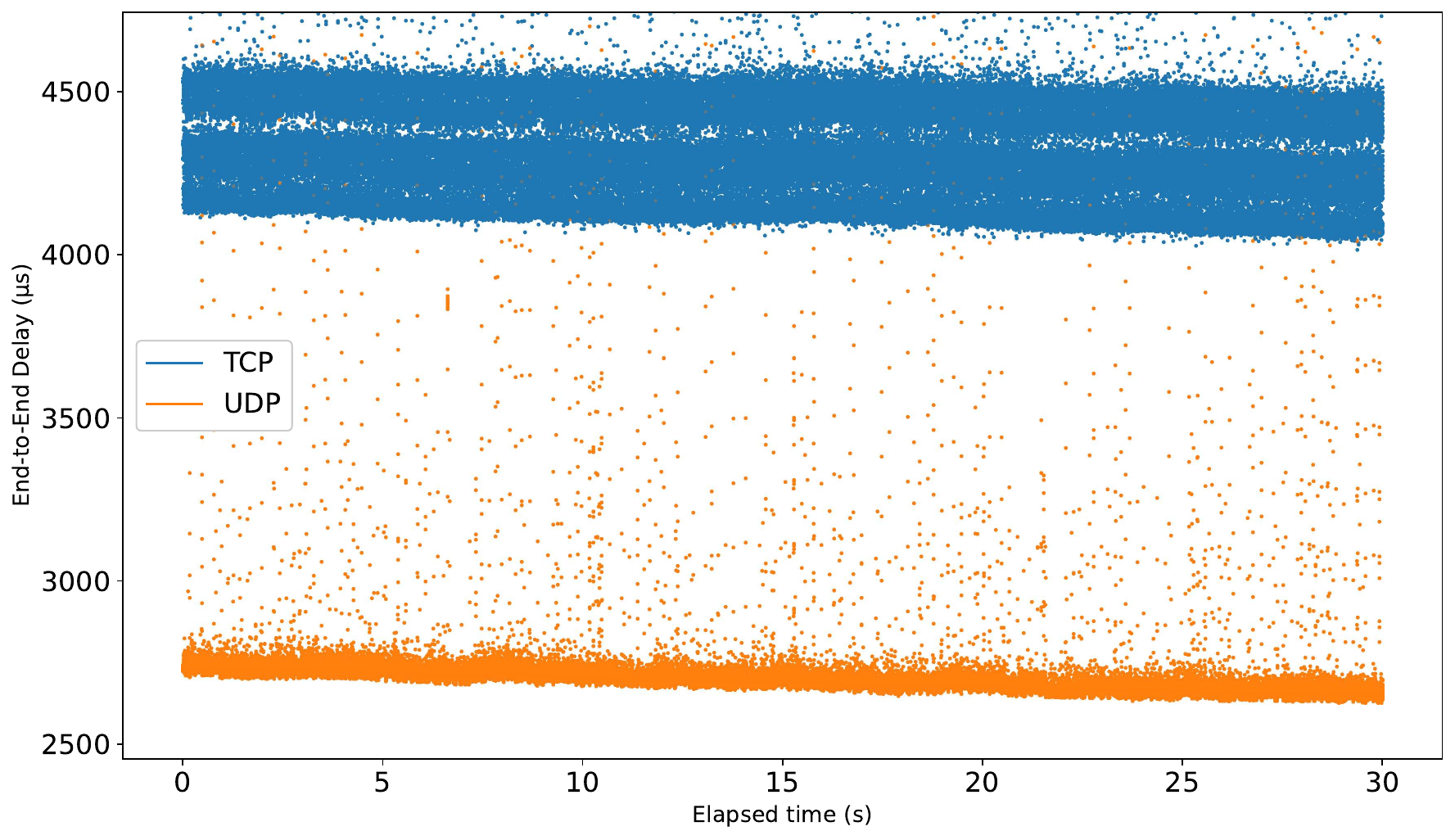} 
        \caption{High-Rate Tests End-to-end Latency.}
        \label{fig:stress_e2e_lat}
\end{figure}

The high-rate experiments increase the transmission rate to evaluate TCP and UDP behavior under a higher message frequency than in the previous configurations. The objective is not to saturate the underlying network, but to assess whether increasing the offered packet rate introduces additional latency, jitter, or queueing effects at the transport and application levels. TCP reaches an E2E delay of approximately 4.1--4.7 ms, with measurements distributed across several discrete bands separated by roughly 100--150 µs. UDP maintains a lower delay, between 2.65 and 2.8 ms, with a slight decrease over the 30 s observation period. The difference between the two protocols remains approximately 1.4--1.9 ms and does not increase over time, indicating no progressive queue buildup under the tested rate (Figure \ref{fig:stress_e2e_lat}).

\begin{figure}[h!]
        \includegraphics[width=\textwidth]{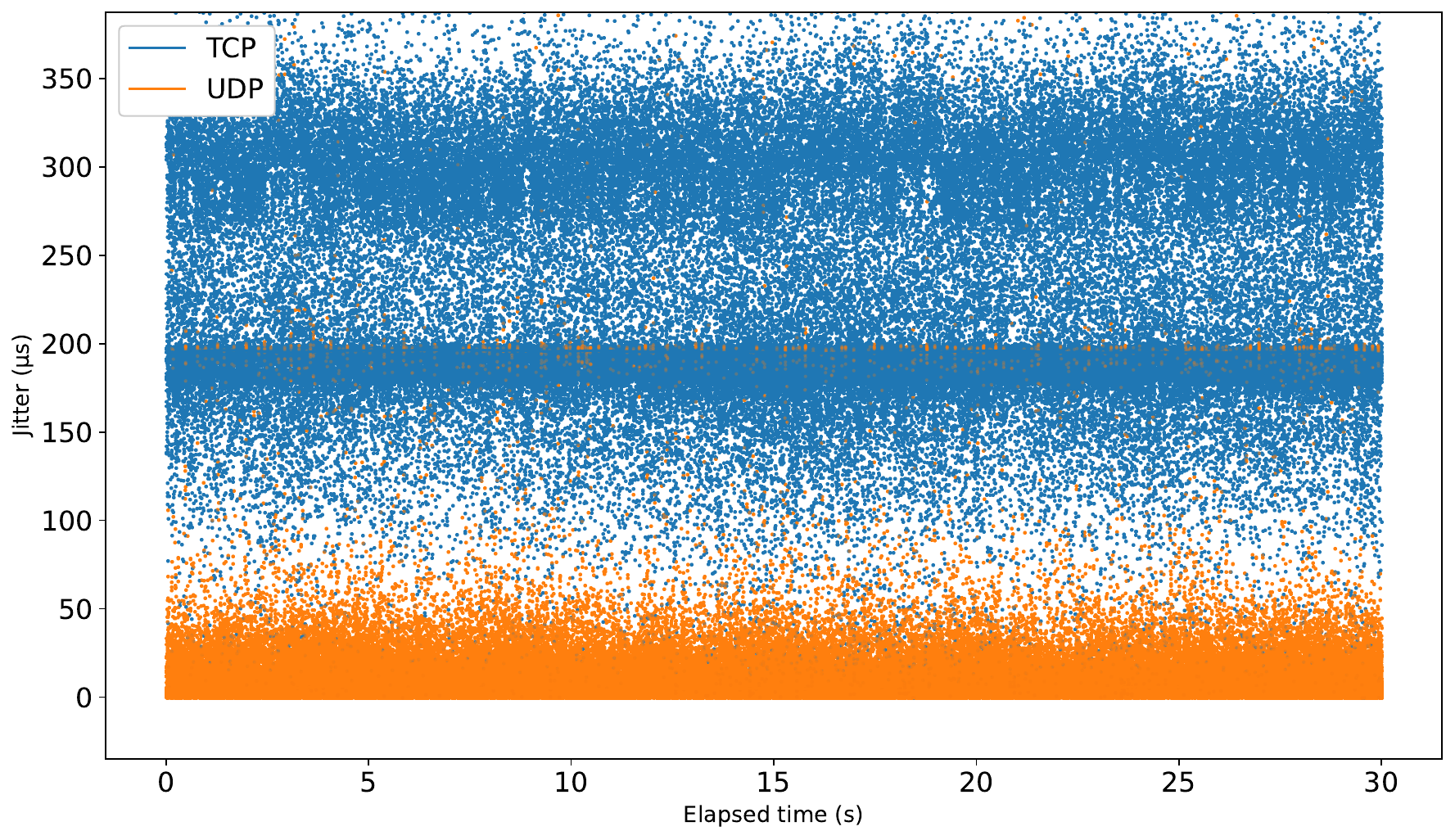} 
        \caption{High-Rate Tests End-to-end Jitter.}
        \label{fig:stress_e2e_jit}
\end{figure}

The jitter results further distinguish the two configurations (Figure \ref{fig:stress_e2e_jit}). TCP shows a bimodal distribution, with a main cluster around 180--200 µs and a tail extending to about 350 µs, whereas UDP remains below 50 µs for the duration of the experiment. Throughput remains stable for both configurations at approximately 2.56 MB/s (20.48 Mbit/s) and 5000 msg/s. All lower-rate configurations achieved a 100\% delivery ratio with no observed message loss. Under the high-rate workload, TCP achieved a delivery ratio of 100\%, while UDP achieved 99.51\%, corresponding to loss rates of 0\% and 0.49\%, respectively. Because this traffic level remains well below the fiber link's capacity, the experiment should be interpreted as a high-message-rate test rather than a network-saturation test. The additional TCP delay and jitter are therefore more likely due to transport-stack and scheduling effects than to link congestion.

\subsection{Delivery Semantics and Load Analysis}
The final experiments examine two operating points: best-effort UDP at 8,000 messages/s, evaluated separately from Section IV-E, and reliability-oriented MQTT QoS 2 at 200 messages/s. Both achieved 100\% delivery with no observed loss. Their different offered loads preclude a direct protocol ranking but illustrate latency under distinct delivery semantics and traffic conditions.

\begin{figure}[h!]
    \includegraphics[width=\textwidth]{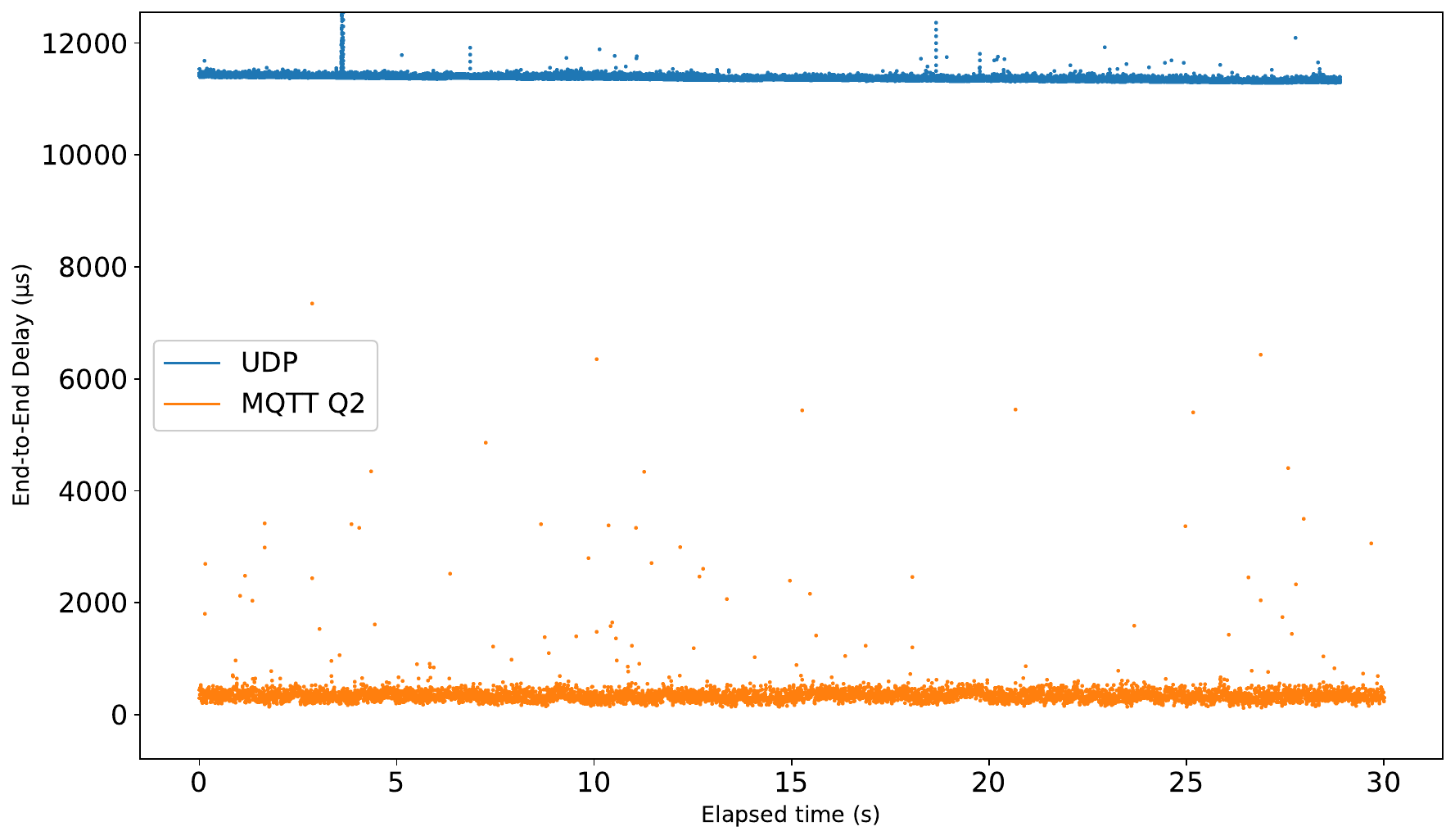}
    \caption{E2E latency in the reliability tests.}
    \label{fig:reliab_e2e_lat}
\end{figure}

UDP E2E delay stabilizes at approximately 11.4 ms, consistent with increased queueing under sustained high-rate traffic despite its low protocol overhead. MQTT QoS 2 achieves a median below 0.5 ms at the lower offered load, despite the PUBLISH/PUBREC/PUBREL/PUBCOMP exchange. This behavior is consistent with limited queueing, although E2E latency does not capture the full QoS 2 handshake duration (Figure \ref{fig:reliab_e2e_lat}). Instantaneous jitter ranges from approximately 0--80~$\mu$s for UDP and 50--75~$\mu$s for MQTT QoS 2.

These results do not establish that MQTT QoS 2 is inherently faster than UDP. They show that low protocol overhead alone does not ensure low latency under high load, while stronger delivery guarantees can coexist with low latency at lower load. Communication performance must therefore be assessed in the context of the pipeline and its operating conditions.

\section{Conclusions}
\label{sec_conclusions}
This paper presented a controlled benchmark of transport, serialization, and compression, evaluated individually and in combination on MASA living-lab hardware. Across the tested configurations, local processing costs alone did not predict communication E2E latency.

Baseline tests showed similarly low delay for UDP and TCP under moderate load, while ZeroMQ exhibited higher delay and jitter at the same throughput. Protobuf reduced serialization time by approximately fourfold compared with CBOR, but communication latency varied independently across configurations. Likewise, the compressed ZeroMQ pipeline achieved lower E2E communication latency than its TCP counterpart, despite higher local compression and decompression costs.

Under the common 30 Hz workload, TCP had the highest median delay, while UDP, ZeroMQ, and both MQTT QoS levels performed similarly. ZeroMQ showed a gradual delay increase consistent with queue accumulation. At higher rates, TCP and UDP maintained a stable delay gap without progressive queue buildup. In the reliability tests, high-rate UDP exhibited substantial queueing, whereas MQTT QoS 2 maintained low latency at a much lower offered load; these results therefore do not establish a latency ranking between the two.

Overall, pipeline performance depends on interactions among transport, middleware, and data processing, with no simple monotonic relationship between reliability guarantees, protocol overhead, and latency. Future work will examine wireless and lossy links and additional protocols, serialization formats, and compression mechanisms.

\balance
\bibliographystyle{IEEEtran}
\bibliography{wm25}

@INPROCEEDINGS{9803729,
  author={Phung, Cao Vien and Bensalem, Mounir and Jukan, Admela},
  booktitle={2022 45th Jubilee International Convention on Information, Communication and Electronic Technology (MIPRO)}, 
  title={Benchmarking Buffer Size in IoT Devices Deploying REST HTTP}, 
  year={2022},
  volume={},
  number={},
  pages={409-414},
  doi={10.23919/MIPRO55190.2022.9803729}}

@ARTICLE{8661513,
  author={Iglesias-Urkia, Markel and Casado-Mansilla, Diego and Mayer, Simon and Bilbao, Josu and Urbieta, Aitor},
  journal={IEEE Internet of Things Journal}, 
  title={Integrating Electrical Substations Within the IoT Using IEC 61850, CoAP, and CBOR}, 
  year={2019},
  volume={6},
  number={5},
  pages={7437-7449},
  doi={10.1109/JIOT.2019.2903344}}

@INPROCEEDINGS{10410639,
  author={Malhi, Avleen and Javed, Asad and Yousefnezhad, Narges and Främling, Kary},
  booktitle={2023 10th International Conference on Future Internet of Things and Cloud (FiCloud)}, 
  title={IoT Open Messaging Standards: Performance Comparison with MQTT and CoAP Protocols}, 
  year={2023},
  volume={},
  number={},
  pages={130-135},
  doi={10.1109/FiCloud58648.2023.00027}}

@INPROCEEDINGS{8248196,
  author={Handosa, Mohamed and Gračanin, Denis and Elmongui, Hicham G.},
  booktitle={2017 Winter Simulation Conference (WSC)}, 
  title={Performance evaluation of MQTT-based internet of things systems}, 
  year={2017},
  volume={},
  number={},
  pages={4544-4545},
  doi={10.1109/WSC.2017.8248196}}

@INPROCEEDINGS{8914552,
  author={Moraes, Thays and Nogueira, Bruno and Lira, Victor and Tavares, Eduardo},
  booktitle={2019 IEEE International Conference on Systems, Man and Cybernetics (SMC)}, 
  title={Performance Comparison of IoT Communication Protocols}, 
  year={2019},
  volume={},
  number={},
  pages={3249-3254},
  doi={10.1109/SMC.2019.8914552}}

@ARTICLE{10734127,
  author={Jackson, Samuel and Cummings, Nathan and Khan, Saiful},
  journal={IEEE Access}, 
  title={Streaming Technologies and Serialization Protocols: Empirical Performance Analysis}, 
  year={2024},
  volume={12},
  number={},
  pages={158025-158039},
  doi={10.1109/ACCESS.2024.3486054}}

@inproceedings{kang2020study,
  title={A study of publish/subscribe middleware under different iot traffic conditions},
  author={Kang, Zhuangwei and Canady, Robert and Dubey, Abhishek and Gokhale, Aniruddha and Shekhar, Shashank and Sedlacek, Matous},
  booktitle={Proceedings of the International Workshop on Middleware and Applications for the Internet of Things},
  pages={7--12},
  year={2020}
}

@inproceedings{proos2020performance,
  title={Performance comparison of messaging protocols and serialization formats for digital twins in IoV},
  author={Proos, Daniel Persson and Carlsson, Niklas},
  booktitle={2020 IFIP networking conference (networking)},
  pages={10--18},
  year={2020},
  organization={IEEE}
}

@inproceedings{gheorghe2020performance,
  title={A performance benchmarking methodology for MQTT broker implementations},
  author={Gheorghe-Pop, Ilie-Daniel and Kaiser, Alexander and Rennoch, Axel and Hackel, Sascha},
  booktitle={2020 IEEE 20th International Conference on Software Quality, Reliability and Security Companion (QRS-C)},
  pages={506--513},
  year={2020},
  organization={IEEE}
}

@INPROCEEDINGS{11096145,
  author={Pietri, Marcello and Taccini, Lorenzo and Bedogni, Luca and Picone, Marco and Mamei, Marco and Zambonelli, Franco},
  booktitle={2025 21st International Conference on Distributed Computing in Smart Systems and the Internet of Things (DCOSS-IoT)}, 
  title={Efficient and Flexibile IoT Communication Through a Plugin-Based MQTT Processing Architecture}, 
  year={2025},
  volume={},
  number={},
  pages={27-34},
  doi={10.1109/DCOSS-IoT65416.2025.00013}}

@INPROCEEDINGS{10770429,
  author={Grazia, Carlo Augusto and Iandolo, Salvatore and Klapez, Martin and Casoni, Maurizio},
  booktitle={2024 20th International Conference on Wireless and Mobile Computing, Networking and Communications (WiMob)}, 
  title={Preliminary test on the Modena Automotive Smart Area Backbone}, 
  year={2024},
  volume={},
  number={},
  pages={74-79},
  doi={10.1109/WiMob61911.2024.10770429}}

\end{document}